\documentclass[prd,aps,12pt,preprint,floatfix,nofootinbib]{revtex4}

\usepackage{amsmath,amssymb}
\usepackage{graphicx}
\usepackage{setspace}
\usepackage{color}

\begin{document}

\title{A qubit model for U(1) lattice gauge theory\footnote{This is the written version of a presentation given at \\
{\em Lattice for Beyond the Standard Model Physics,} \\
Syracuse University, USA, May 2-3, 2019. \\
\url{http://www-hep.colorado.edu/~eneil/lbsm19}}}

\author{Randy Lewis$^1$ and R. M. Woloshyn$^2$}
\affiliation{$^1$Department of Physics and Astronomy, York University, Toronto, ON, M3J 1P3, Canada \\
             $^2$TRIUMF, 4004 Wesbrook Mall, Vancouver, BC, V6T 2A3, Canada \\
             \linepenalty=300}

\begin{abstract}
A conceptually simple model for strongly interacting compact U(1) lattice gauge theory is expressed as operators acting on qubits.
The number of independent gauge links is reduced to its minimum through the use of Gauss's law.
The model can be implemented with any number of qubits per gauge link, and a choice as small as two is shown to be useful.
Real-time propagation and real-time collisions are observed on lattices in two spatial dimensions.
The extension to three spatial dimensions is also developed, and a first look at 3-dimensional real-time dynamics is
presented.
\end{abstract}

\maketitle

\section{Motivation}

Numerical simulations of lattice gauge theory started about 40 years ago \cite{Creutz:1979dw}.
During this time, lattice gauge theory has become the central tool for first-principles studies of nonperturbative quantum chromodynamics (QCD), and therefore it plays a vital role in the field of particle physics \cite{Tanabashi:2018oca}.
Lattice gauge theory is also being applied vigorously to theories beyond QCD \cite{Brower:2019oor}.
For QCD and beyond, the standard approach has been Monte Carlo calculations in Euclidean spacetime, which do not provide access to real-time dynamics.
Avoiding this Monte Carlo approach is difficult because the required Hilbert space is usually far larger than any classical computer can hope to manage \cite{Preskill:2018fag}.

The availability of a practical quantum computer would offer a different approach to lattice gauge theory.
Calculations could be performed for real time, allowing access to the dynamics that are inaccessible to classical computers.
Doubling the size of the Hilbert space would only require the addition of a single qubit.
Recent studies have explored various questions related to the quantum computation of lattice gauge theories \cite{Osterloh:2005ha,Buchler:2005fq,Byrnes:2005qx,Zohar:2012ay,Zohar:2012xf,Banerjee:2012pg,Banerjee:2012xg,Wiese:2013uua,Bazavov:2015kka,Zohar:2016iic,Bermudez:2017yrq,Banuls:2017ena,Kaplan:2018vnj,Stryker:2018efp,Kokail:2018eiw,Raychowdhury:2018osk,Zhang:2018ufj,Hackett:2018cel,Lamm:2019bik}.

This paper describes a qubit model for the simplest lattice gauge theory: compact U(1) in the absence of charged particles.
Recall that compactness ensures that this theory is not a trivial theory of free photons, but has a strong coupling region where the theory has confinement and particles akin to glueballs
\cite{Polyakov:1976fu,Hasenfratz:1980ue,Irving:1983ms,Yung:1988jr,Hamer:1991dy,Chaara:1994ar,Panero:2005iu,Athenodorou:2018sab}.
Our model is meant to represent compact U(1) in the strong coupling region.
The model is obtained by first writing the Hamiltonian in terms of a minimal set of independent gauge links in a manner that maintains Gauss's law explicitly, then choosing a particular set of basis states so the Hamiltonian is expressed in terms of Pauli operators that would act on a register of qubits, and then truncating that basis set to a particular subset.
Through direct calculations, the truncation error is observed to be small and manageable.

Section \ref{sec:oneplaq} describes the model on the smallest lattice: a single plaquette (four sites in the shape of a square).
Section \ref{sec:rowplaq} uses the model on a 1-dimensional row of plaquettes to show the dynamics of excitations.
We find that real-time propagation and real-time collisions can be observed quite readily in this model.
Section \ref{sec:2D} applies the model to a 2-dimensional lattice of plaquettes.
Section \ref{sec:3D} explains a new issue that arises when extending the model to 3-dimensional lattices, and how it can be handled.
A brief conclusion is given in Section \ref{sec:concl}.
All of the calculations for this work were performed using simulators running on classical computers.

\section{The model on one plaquette}\label{sec:oneplaq}

The Euclidean lattice action for U(1) gauge theory is well known \cite{Kogut:1979wt}.
For any hypercubic lattice, it is a sum over all plaquettes,
\begin{eqnarray}
S &=& -\frac{\beta}{2}\sum_P\bigg(U_P+U_P^*\bigg) \,, \\
U_P &=& e^{i\theta_\mu(n)}e^{i\theta_\nu(n+\hat\mu)}e^{-i\theta_\mu(n+\hat\nu)}e^{-i\theta_\nu(n)} \,,
\end{eqnarray}
where $\beta=1/g^2$ and $\theta_\mu(n)=agA_\mu(n)$, with $g$ the bare gauge coupling and $A_\mu(n)$ the vector potential.
Choosing the temporal gauge (meaning $A_t(n)=0$) allows the continuum limit to be taken in the temporal direction,
and brings us to the Hamiltonian for compact U(1) gauge theory:
\begin{equation}
H = \sum_n\left(\frac{a^3}{2}\vec E^2(n)
    -\frac{\beta}{a}\sum_{i=2}^3\sum_{j=1}^{i-1}\cos\bigg(\theta_i(n)+\theta_j(n+\hat i)-\theta_i(n+\hat j)-\theta_j(n)\bigg)\right)
\end{equation}
where $\vec E = \partial_t\vec A$ is the electric field.
The conjugate momentum for the angle $\vec\theta$ (because it is a kind of angular momentum, let's name it $\vec L$) is proportional to $\vec E$, and obeys the canonical commutation relation,
\begin{equation}
[\theta_j(n),L_k(n^\prime)] = i\delta_{jk}\delta_{nn^\prime} \,.
\end{equation}
This implies that link variables, $U_j=e^{i\theta_j}$, are ladder operators because they obey the corresponding commutation relations,
\begin{eqnarray}
\left[U_j,L_j\right] &=& -U_j \,, \\
\left[U_j^\dagger,L_j\right] &=& U_j^\dagger \,.
\end{eqnarray}
Following \cite{Byrnes:2005qx}, we can define a state $\left|0\right>$ by
\begin{equation}
L_j\left|0\right> = 0
\end{equation}
and then identify a sequence of states with integer eigenvalues,
\begin{eqnarray}
L_j(U_j)^\ell\left|0\right> &=& \ell \, (U_j)^\ell\left|0\right> \,, \\
L_j(U_j^\dagger)^\ell\left|0\right> &=& -\ell \, (U_j^\dagger)^\ell\left|0\right> \,,
\end{eqnarray}
where $\ell = \ldots, -2, -1, 0, 1, 2, \ldots$
The Hamiltonian can now be written as
\begin{equation}\label{eq:generalH}
H = \frac{1}{2a\beta}\bigg(\sum_jL_j^2 - \beta^2\sum_P(U_P+U_P^\dagger)\bigg) \,.
\end{equation}
Let's rescale energies by dropping the overall factor of $(2a\beta)^{-1}$ from now on.
Having arrived at a convenient form for the Hamiltonian of compact U(1) lattice gauge theory, we now apply it to the smallest possible lattice.

\begin{figure}[tb]
\includegraphics[scale=0.5,clip=true,trim=0 40 0 80]{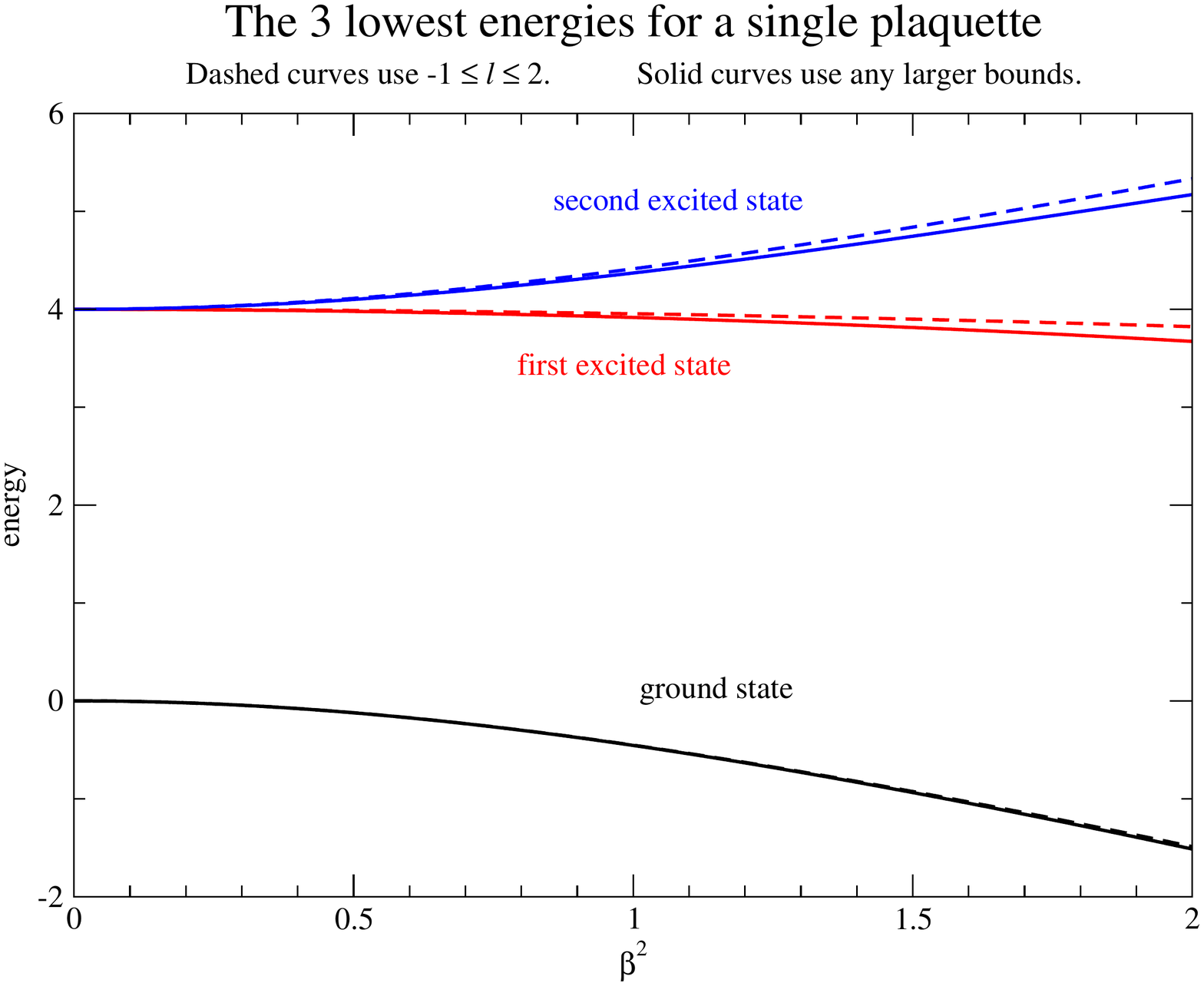}
\caption{The three lowest eigenvalues of the Hamiltonian matrix for a single plaquette.
Dashed curves use $-1\leq\ell\leq2$.
Solid curves use any larger bounds.}
\label{fig:plaqLowStates}
\end{figure}

Consider a 2$\times$2 lattice with Dirichlet boundary conditions.
This lattice has four sites, four gauge links and only one plaquette so
\begin{equation}
H = L_1^2 + L_2^2 + L_3^2 + L_4^2 - \beta^2(U_1U_2U_3^\dagger U_4^\dagger+U_1^\dagger U_2^\dagger U_3U_4) \,.
\end{equation}
Because there are no charges in a pure gauge theory, Gauss's law ensures that electric flux is conserved at every lattice site so $L_i^2$ is the same for all four links.
Moreover, a local gauge transformation at any lattice site allows one adjoining link to be rotated to the identity, and this can be done at three sites on our lattice, leaving a minimal Hamiltonian:
\begin{equation}\label{eq:oneplaqH}
H = 4L^2 - \beta^2(L^++L^-)
\end{equation}
where $L^\pm$ are the ladder operators.
Evaluating the Hamiltonian in the basis of electric field eigenstates and truncating to $-4\leq\ell\leq4$ yields the Hamiltonian matrix
\begin{equation}\label{eq:Hmatrix}
\left(\begin{array}{ccccccccc}
64 & -\beta^2 & 0 & 0 & 0 & 0 & 0 & 0 & 0 \\
-\beta^2 & 36 & -\beta^2 & 0 & 0 & 0 & 0 & 0 & 0 \\
0 & -\beta^2 & 16 & -\beta^2 & 0 & 0 & 0 & 0 & 0 \\
0 & 0 & -\beta^2 & 4 & -\beta^2 & 0 & 0 & 0 & 0 \\
0 & 0 & 0 & -\beta^2 & 0 & -\beta^2 & 0 & 0 & 0 \\
0 & 0 & 0 & 0 & -\beta^2 & 4 & -\beta^2 & 0 & 0 \\
0 & 0 & 0 & 0 & 0 & -\beta^2 & 16 & -\beta^2 & 0 \\
0 & 0 & 0 & 0 & 0 & 0 & -\beta^2 & 36 & -\beta^2 \\
0 & 0 & 0 & 0 & 0 & 0 & 0 & -\beta^2 & 64 \\
\end{array}\right) \,.
\end{equation}
Because the on-diagonal entries grow quadratically while the off-diagonals remain constant as we approach the corners of the matrix, truncation has a small effect on the lowest eigenvalues at strong coupling (i.e.\ small $\beta^2$).
To be specific, in Fig.~\ref{fig:plaqLowStates} it was necessary to truncate to a 4$\times$4 Hamiltonian in order to display a visible deviation, and even then the ground state deviation is barely seen.
Notice that the 4$\times$4 case corresponds to non-symmetric limits because $\ell=-1,0,1,2$.

To run someday on a quantum computer (and today on a classical simulator),
the states have to be mapped to qubits, and operators have to be expressed as products of Pauli operators
acting on those qubits.
A proof of principle was provided in  \cite{Byrnes:2005qx} though the authors acknowledged that it was a qubit-inefficient implementation.
A more efficient alternative here will be used here.

An optimal description of the $\ell=-1,0,1,2$ truncation needs only two qubits.
Including a third qubit doubles the number of basis states.
Our mapping conventions for those two cases are displayed in Tab.~\ref{tab:states}.
\begin{table}[tb]
\caption{The correspondence between electric field eigenvalues (named $\ell$) and states of the qubit register for two cases: 2 qubits and 3 qubits.  When referenced in the text, qubits within a register are numbered from right to left: \ldots210.}
\begin{tabular*}{5cm}{@{\extracolsep{\stretch{1}}}*{4}{r}@{}}
\hline\hline
\multicolumn{2}{c}{2 qubits} & \multicolumn{2}{c}{3 qubits} \\
\hline
$\ell$ & state & $\ell$ & state \\
\hline
 2 & $\left|00\right>$ & 4 & $\left|000\right>$ \\
 1 & $\left|01\right>$ & 3 & $\left|001\right>$ \\
 0 & $\left|10\right>$ & 2 & $\left|010\right>$ \\
-1 & $\left|11\right>$ & 1 & $\left|011\right>$ \\
   &                   & 0 & $\left|100\right>$ \\
   &                  & -1 & $\left|101\right>$ \\
   &                  & -2 & $\left|110\right>$ \\
   &                  & -3 & $\left|111\right>$ \\
\hline\hline
\end{tabular*}
\label{tab:states}
\end{table}
The operators that appear in the Hamiltonian, Eq.~(\ref{eq:oneplaqH}), can be written as follows:
\begin{eqnarray}
L^+ &=& \sigma_0^+ + \sigma_0^-\sigma_1^+ + \sigma_0^-\sigma_1^-\sigma_2^+ \,, \label{eq:Lplus} \\
L^- &=& \sigma_0^- + \sigma_0^+\sigma_1^- + \sigma_0^+\sigma_1^+\sigma_2^- \,, \\
L &=& \tfrac{1}{2}\left(1+\sigma_0^z+2\sigma_1^z+4\sigma_2^z\right) \,, \label{eq:L}
\end{eqnarray}
where $\sigma_i$ acts on the $i$th qubit in the register.
All terms containing $\sigma_2$ can simply be omitted for the case of a two-qubit register.
Conversely, an extension to a register with more than three qubits is straightforward though unnecessary for our purposes.

Translating from the qubit ladder operators $\sigma^\pm$ to standard Pauli operators, the two-qubit Hamiltonian is
\begin{equation}
H = 6 + 2\sigma^z_0 + 4\sigma^z_1 + 4\sigma^z_0\sigma^z_1
      - \beta^2\left(\sigma^x_0 + \frac{1}{2}(\sigma^x_0\sigma^x_1+\sigma^y_0\sigma^y_1)\right)
\end{equation}
and the three-qubit Hamiltonian is
\begin{eqnarray}
H &=& 22 + 2\sigma^z_0 + 4\sigma^z_1 + 8\sigma^z_2
      + 4\sigma^z_0\sigma^z_1 + 8\sigma^z_0\sigma^z_2 + 16\sigma^z_1\sigma^z_2 \nonumber \\
   && -\beta^2\left(\sigma^x_0 + \frac{1}{2}(\sigma^x_0\sigma^x_1+\sigma^y_0\sigma^y_1)
                + \frac{1}{4}(\sigma^x_0\sigma^x_1\sigma^x_2-\sigma^y_0\sigma^y_1\sigma^x_2
                             +\sigma^y_0\sigma^x_1\sigma^y_2+\sigma^x_0\sigma^y_1\sigma^y_2)\right) \,.
\end{eqnarray}
The time evolution operator in a quantum theory is $e^{-iHt}$ which requires exponentiating the Hamiltonian.
The second-order Suzuki-Trotter formula \cite{Lloyd},
\begin{equation}\label{trotter}
e^{-i(A+B)t} = e^{-iAt/2}e^{-iBt}e^{-iAt/2} + O(t^3) \,,
\end{equation}
is used to rewrite the time evolution, which is due to sums of non-commuting terms in the Hamiltonian, into products of unitary operators.
High level quantum computer programming languages being developed today provide for exponentiated operators such as $e^{i\sigma^jt}$ for $j=x,y,z$ .

\begin{figure}[tb]
\includegraphics[scale=0.5,clip=true,trim=0 40 0 80]{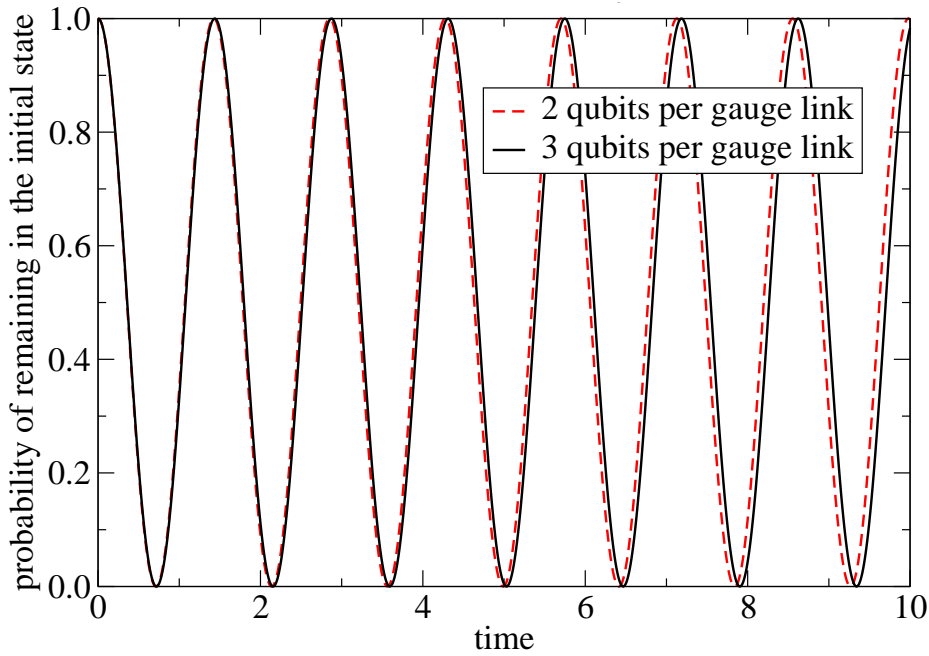}
\caption{Time evolution of the one-plaquette Hamiltonian using the two-qubit model and the three-qubit model.}
\label{fig:superposition}
\end{figure}

As an explicit example of time evolution, choose $\beta^2=1$ and begin with an initial state (at $t=0$) that is equally weighted between the ground and first excited states,
\begin{equation}
\left|{\rm init}\right> = \frac{1}{\sqrt{2}}\left|E_0\right> + \frac{1}{\sqrt{2}}\left|E_1\right> \,.
\end{equation}
As shown in Fig.~\ref{fig:superposition}, the probability of remaining in the initial state oscillates between zero and unity.
Several oscillations pass before a distinction is noticed between the calculations with two-qubit and three-qubit registers.

For a final calculation on the one-plaquette lattice, we solve for the ground state eigenvalue and eigenstate.
Of course these can be found easily by matrix diagonalization (recall Eq.~(\ref{eq:Hmatrix}) and Fig.~\ref{fig:plaqLowStates}),
but now an implementation of the variational principle for a quantum computer \cite{Peruzzo} will be used.
To begin, write the two-qubit Hamiltonian's expectation value as a function of probabilities,
\begin{eqnarray}
\left<\psi|H|\psi\right> &=& 4 + 12P_{00} - 4P_{10}
-\beta^2(2P^x_{00}+P^x_{10}
-P^{xx}_{01}-P^{xx}_{10}
-P^{yy}_{01}-P^{yy}_{10}) \,, \\
P_i &=& \left|\left<i|\psi\right>\right|^2 \,, \label{eq:prob1} \\
P_i^x &=& \left|\left<i|R^y(-\tfrac{\pi}{2},0)\psi\right>\right|^2 \,, \\
P_i^{xx} &=& \left|\left<i|R^y(-\tfrac{\pi}{2},0)R^y(-\tfrac{\pi}{2},1)\psi\right>\right|^2 \,, \\
P_i^{yy} &=& \left|\left<i|R^x(\tfrac{\pi}{2},0)R^x(\tfrac{\pi}{2},1)\psi\right>\right|^2 \,. \label{eq:prob4}
\end{eqnarray}
The quantum computer can be repeatedly prepared in state $\left|\psi\right>$, and then measurements of the
appropriate spin states will determine the probabilities in Eqs.~(\ref{eq:prob1}-\ref{eq:prob4}).
Those, in turn, give the Hamiltonian's expectation value.

For a simple trial state, we anticipate that the ground state will be dominated by $\ell=0$, that it will have essentially equal contributions from $\ell=\pm1$, and that it will have a negligible contribution from $\ell=2$.
Therefore we use
\begin{equation}\label{eq:trial}
\left|\psi\right> = 0\left|00\right>
+ \tfrac{1}{\sqrt{2}}\sin\theta\left|01\right>
+\cos\theta\left|10\right>
+\tfrac{1}{\sqrt{2}}\sin\theta\left|11\right> \,.
\end{equation}
A four-step procedure will create this state in the quantum register.
Step~1:~Begin with $\left|00\right>$,
Step~2:~Apply X(1) to get $\left|10\right>$,
Step~3:~Apply $R^y(2\theta,0)$ to get
$\cos\theta\left|10\right>+\sin\theta\left|11\right>$,
Step~4:~Apply $CR^y(-\tfrac{\pi}{2},0,1)$ to get $\left|\psi\right>$.

\begin{figure}[tb]
\includegraphics[scale=0.5,clip=true,trim=0 35 0 80]{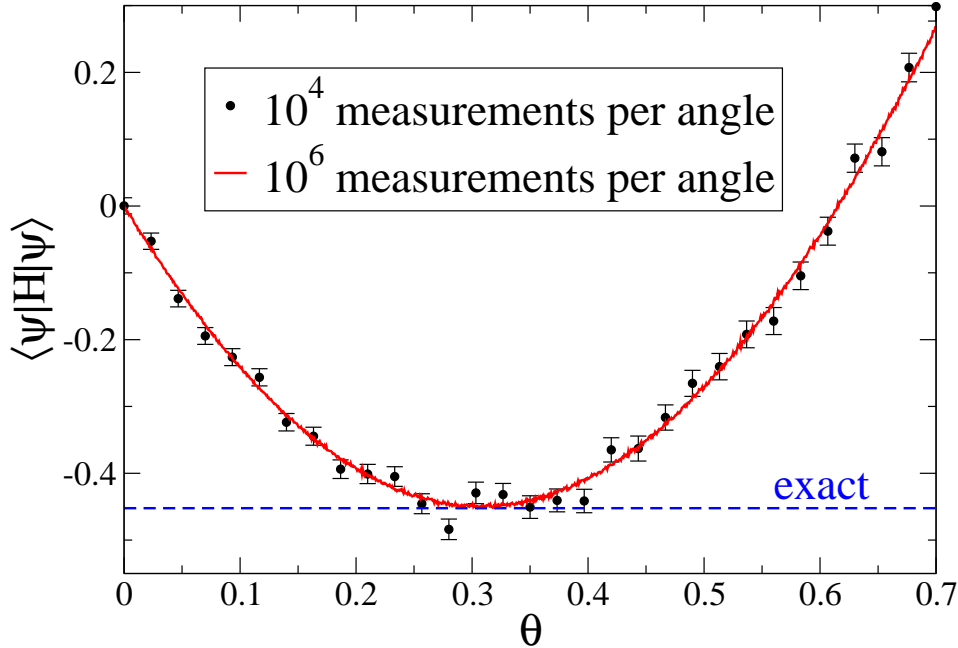}
\caption{Using the quantum variational principle to determine the ground state eigenvalue for the one-plaquette model at $\beta^2=1$.
The variational parameter $\theta$ is defined in Eq.~(\ref{eq:trial}).
Error bars are quantum measurement uncertainties.
The value labeled {\em exact} is from matrix diagonalization.}
\label{fig:vqe}
\end{figure}

A calculation of the Hamiltonian's expectation value is plotted in Fig.~\ref{fig:vqe}.
Our simple trial state identifies an optimal value for $\theta$ that gives a good estimate of the exact ground state eigenvalue.
As the plot indicates, many measurements are needed before a precise determination is obtained.

\section{A one-dimensional array of plaquettes}\label{sec:rowplaq}

To explore the real-time propagation of excitations, we now extend the lattice from a single plaquette to a row of plaquettes.
The Hamiltonian is again Eq.~(\ref{eq:generalH}) and the factor of $(2a\beta)^{-1}$ can again be dropped.
Application of Gauss's law and gauge invariance reduces the Hamiltonian to one independent gauge link per plaquette.
For example, a row of three plaquettes gives
\begin{eqnarray}
H &=& 3L_1^2+2L_2^2+3L_3^2+(L_2-L_1)^2+(L_3-L_2)^2-\beta^2(L_1^++L_1^-+L_2^++L_2^-+L_3^++L_3^-) \nonumber \\
  &=& 4L_1^2+4L_2^2+4L_3^2-2L_1L_2-2L_2L_3-\beta^2(L_1^++L_1^-+L_2^++L_2^-+L_3^++L_3^-) \label{eq:Hrow}
\end{eqnarray}
which is simply three copies of the single-plaquette Hamiltonian plus an extra term for each gauge link that is shared between plaquettes.
The operators for each of the three links are given by Eqs.~(\ref{eq:Lplus}-\ref{eq:L}).
The extension to any number of plaquettes in a row is straightforward.

For definiteness, choose a row of seven plaquettes described by two qubits per gauge link.
The time evolution of that system can be computed on a 14-qubit register from any chosen initial state.
Consider the initial state where the first plaquette begins in its first excited state and all others begin in the ground state.
Time evolution will allow us to observe the motion of that excitation in real time.
The exact eigenvalues and eigenstates for an individual plaquette can be determined by diagonalizing
the appropriate 4$\times$4 truncation of Eq.~(\ref{eq:Hmatrix}), and that is what will be used to build our initial state.

\begin{figure}[tb]
\includegraphics[scale=0.5,clip=true,trim=0 40 0 80]{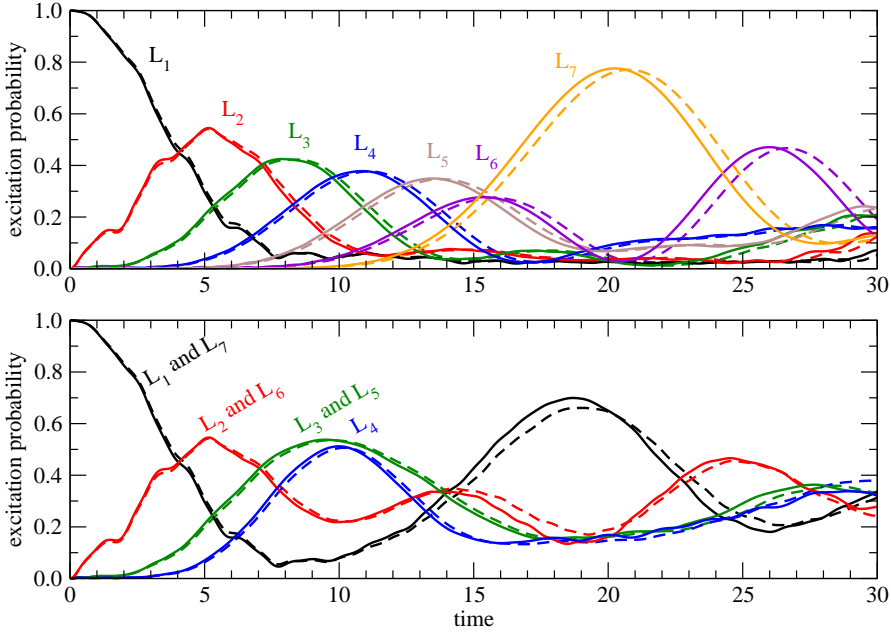}
\caption{Time evolution on a 2$\times$8 lattice at $\beta^2=1$ using 2 qubits (dashed) or 3 qubits (solid) per gauge link.
The initial state has the link at one end (upper plot) or the links at both ends (lower plot) in the single-plaquette first excited state and all others in the single-plaquette ground state.}
\label{fig:row7}
\end{figure}

The upper plot in Fig.~\ref{fig:row7} shows our results at $\beta^2=1$, calculated by expressing Eq.~(\ref{eq:Hrow})
in terms of Pauli operators and then using the second-order Trotter formula, Eq.~(\ref{trotter}), to translate into
unitary operators acting on the qubit register.
Most calculations were performed on a homemade simulator written in fortran, with some checks carried out in the wavefunction simulators of pyQuil \cite{Smith} and ProjectQ \cite{Steiger}.
The excitation that begins at $L_1$ moves across the lattice to $L_7$ and then bounces back.
Only a modest dependence on the number of qubits is observed.
The small jagged effects in the $L_1$ and $L_2$ curves are reminders that neighbouring plaquettes share a gauge link at
their boundary, leading to the mixing terms in the Hamiltonian of Eq.~(\ref{eq:Hrow}).
This is no problem, it just means that our chosen initial state reflects the fact that the plaquettes on our lattice do not factorize.
Another consequence is that the seven curves in Fig.~\ref{fig:row7} do not add exactly to unity, because the probabilities defined by these curves are not exactly mutually exclusive.
The deviation from unity is roughly 5 percent and, as must be the case, the sum is larger (not smaller) than unity at all times.

The lower plot in Fig.~\ref{fig:row7} shows the time evolution of a different initial state.
In this case, $L_1$ and $L_7$ (at opposite ends of the lattice) both begin in the single-plaquette first excited state while
the others ($L_2$ through $L_6$) begin in the single-plaquette ground state.
Time evolution shows the excitations traveling to meet at the centre of the lattice and then traveling to the lattice walls where they bounce toward one another again.
Comparison of the upper and lower plots indicates that the interaction between two traveling excitations has caused them to traverse the lattice in less time than a single excitation, since the large peak for $L_7$ has moved from time $\gtrsim$20 (upper plot) to time $\lesssim$19 (lower plot).
An attractive interaction between the two excitations would cause this effect, while a repulsive interaction would cause the opposite effect.
The classical analogy is two balls rolling on side-by-side tracks, where one track has a valley and the other has a hill.
Both balls begin and end at the same height and same speed, but the ball with the valley (attractive potential) takes less time and therefore wins the race.

\begin{figure}[tb]
\includegraphics[scale=0.5,clip=true,trim=0 40 0 80]{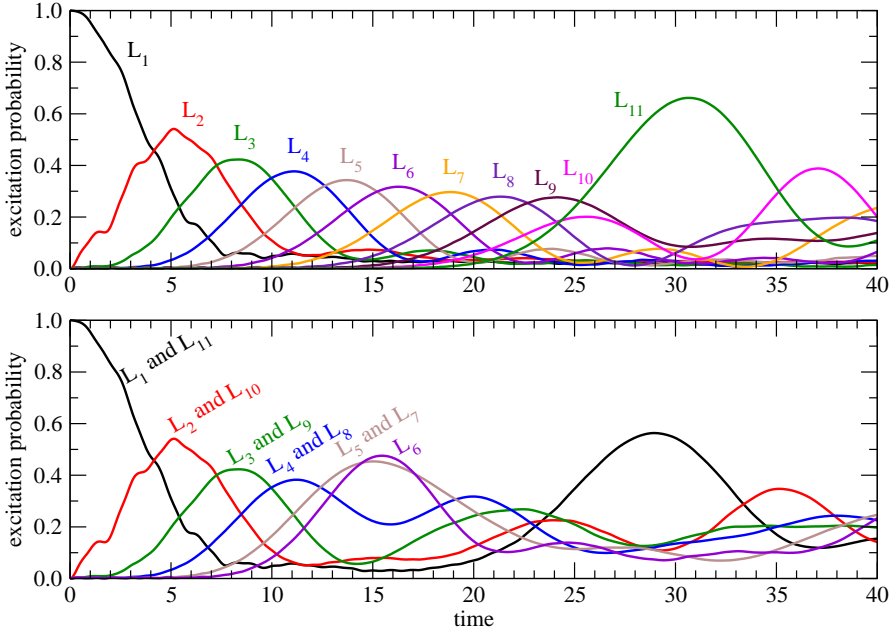}
\caption{Same a Fig.~\ref{fig:row7} but on a 2$\times$12 lattice with 2 qubits per gauge link.}
\label{fig:row11}
\end{figure}

\begin{figure}[tb]
\includegraphics[scale=0.5,clip=true,trim=0 40 0 80]{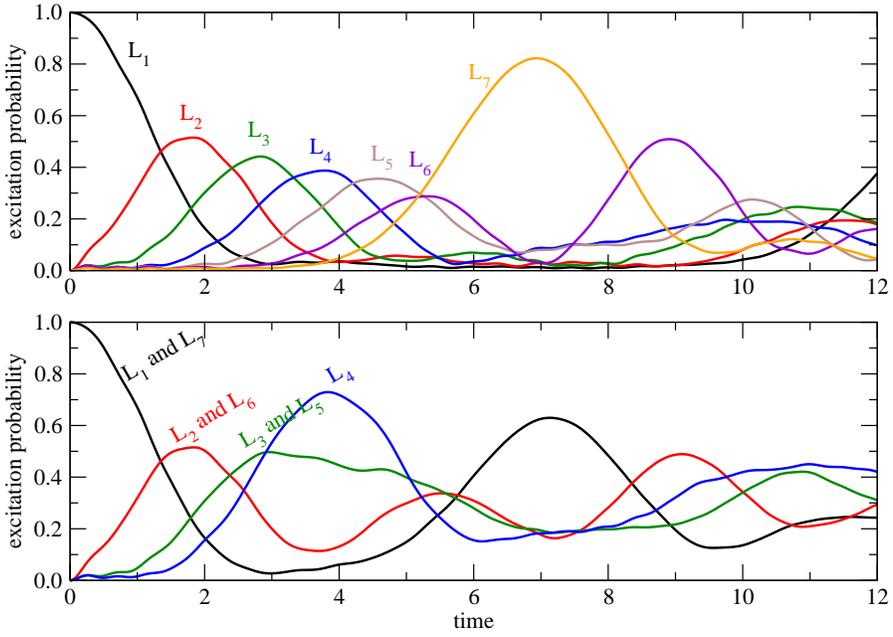}
\caption{Same a Fig.~\ref{fig:row7} but for $\beta^2=3$ with 2 qubits per gauge link.}
\label{fig:row7beta3}
\end{figure}

\begin{figure}[tb]
\includegraphics[scale=0.5,clip=true,trim=0 40 0 80]{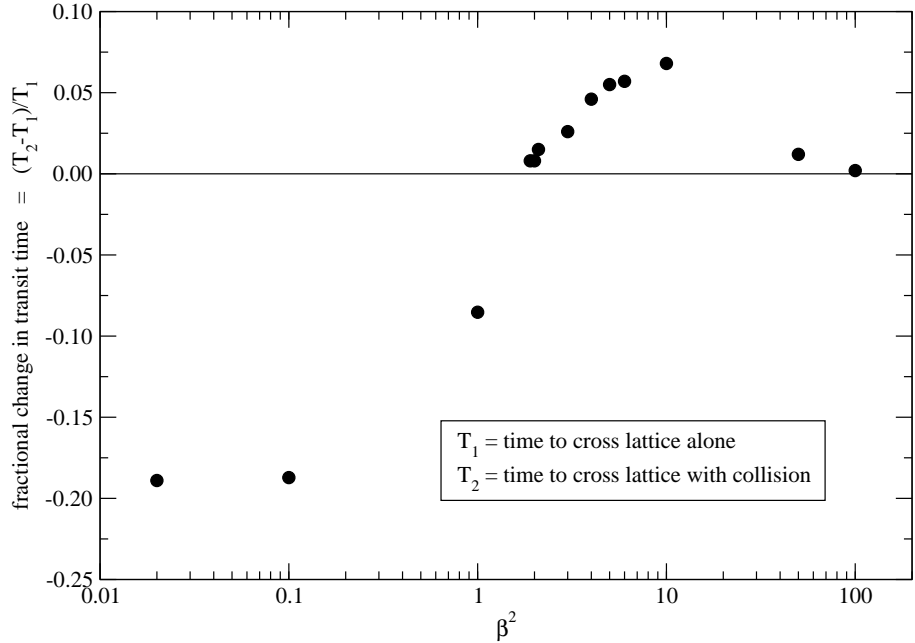}
\caption{The fractional change in transit time for an excitation traversing the lattice alone or with an oncoming excitation.
Calculations are performed on 2$\times$8 lattices using 2 qubits.
The data points are simple estimates of peak locations on plots like Figs.~\ref{fig:row7} and \ref{fig:row7beta3} so no error bars are shown.}
\label{fig:attractrepel}
\end{figure}

Figure~\ref{fig:row11} indicates that the attractive interaction persists on a longer lattice,
while Fig.~\ref{fig:row7beta3} shows a slightly repulsive interaction for a larger value of $\beta^2$.
Exploration of a few other $\beta^2$ values led to Fig.~\ref{fig:attractrepel} which suggests an attractive region at strong coupling and
a repulsive region at weaker coupling.
As the gauge coupling vanishes ($\beta^2\to\infty$), the interaction vanishes.

\section{A two-dimensional lattice}\label{sec:2D}

The model presented in this work is directly applicable to any planar lattice.
As was true for a one-dimensional array of plaquettes, Gauss's law ensures that any two-dimensional array of plaquettes has one independent gauge link per plaquette.
Consider the example of a 4$\times$4 lattice, which is nine plaquettes in the shape of a square.
According to Eq.~(\ref{eq:generalH}), the Hamiltonian is
\begin{eqnarray}
H &=& \sum_{i=1}^9\bigg(4L_i^2-\beta^2(L_i^++L_i^-)\bigg) - 2\sum_{i=1}^6L_iL_{i+3} \nonumber \\
  && - 2(L_1L_2+L_2L_3+L_4L_5+L_5L_6+L_7L_8+L_8L_9) \,.
\end{eqnarray}
After expressing this Hamiltonian in terms of Pauli operators and applying the Trotter formula, Eq.~(\ref{trotter}),
to a chosen initial state, its time evolution is obtained.

\begin{figure}[tb]
\includegraphics[scale=1.0,clip=true,trim=60 20 80 93]{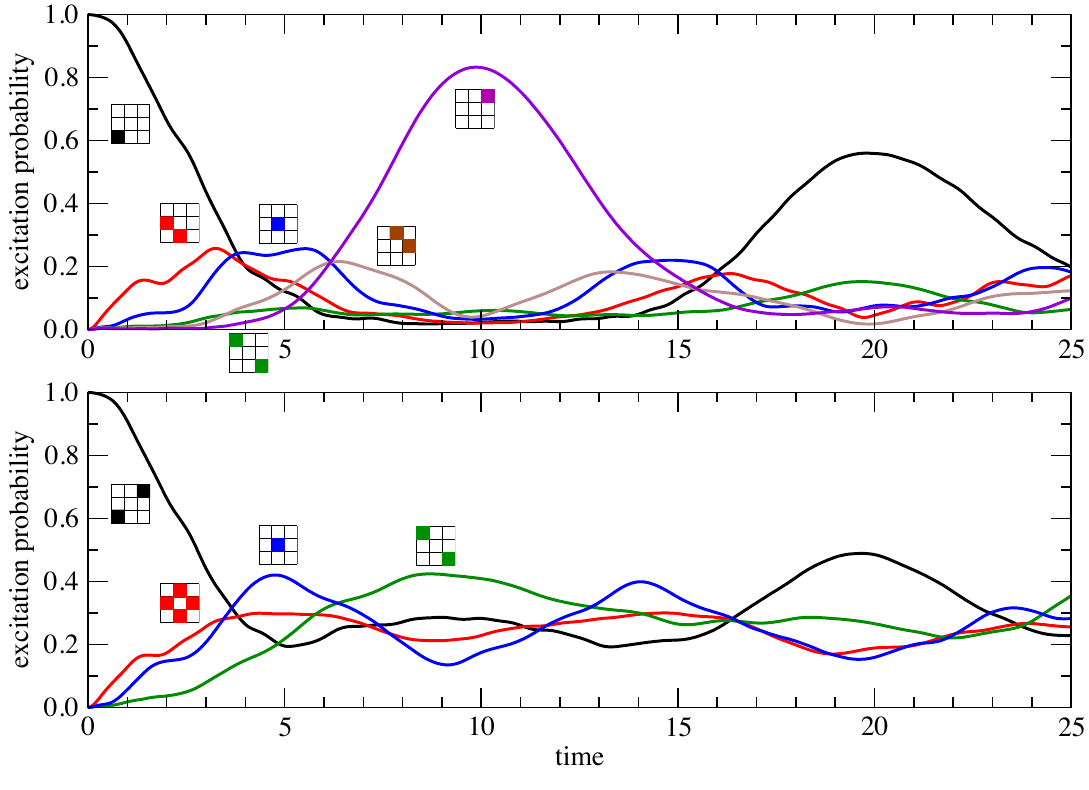}
\caption{Time evolution on a 4$\times$4 lattice at $\beta^2=1$ using 2 qubits per gauge link.
The initial state has the link at one corner (upper plot) or the links at opposite corners (lower plot)
in the single-plaquette first excited state and all others in the single-plaquette ground state.}
\label{fig:squareplots}
\end{figure}
Fig.~\ref{fig:squareplots} shows the case of $\beta^2=1$ for two qubits per gauge link (giving 18 qubits in total).
The upper plot begins with $L_1$ in its first excited state and all others in the ground state, so it corresponds
approximately to one excitation beginning in a corner of the lattice.
We say ``approximately'' because $L_1$ affects the boundaries of neighbouring plaquettes as well.
This was an approximation for the row of plaquettes in Sec.~\ref{sec:rowplaq} also,
but it is more significant now because this square lattice has more boundaries per plaquette than were present in the row of plaquettes.
A qualitative interpretation is clear from the graph.
The excitation travels diagonally across the lattice and then bounces back to the original corner.
The probability of exciting the intermediate corners en route is noticeably smaller than the probability of exciting the center plaquette.

The lower graph in Fig.~\ref{fig:squareplots} shows the time evolution of an initial state where $L_1$ and $L_9$ begin in the
first excited state with all others in the ground state.
Qualitatively, the graph suggests that the two excitations travel from their opposing corners, meet in the center of the lattice,
travel to the corners that have not yet been visited, return to the center, and then return to their original corners.
This particular sequence of events depends on the initial condition and on the choice of $\beta^2$.
The lower graph in Fig.~\ref{fig:squareplots} does not have any peaks that reach above 0.5 after the initial peak,
so it would be challenging to analyze this time evolution from probabilities measured on a quantum computer.

\section{A three-dimensional lattice}\label{sec:3D}

In contrast to the lower-dimensional cases, a cubic spatial lattice will have less than one independent gauge link
per plaquette, and that fact is a significant issue for implementation of the model.
The smallest example is the $2^3$ lattice, which is a single cube and therefore has six plaquettes, but Gauss's law
says that only five gauge links are independent.
To prevent the magnetic part of the Hamiltonian (i.e.\ the terms proportional to $\beta^2$) from ballooning into a huge number
of Pauli-operator terms, it is important to choose the independent links in a clever way.
Additionally, it is important to choose the independent links in such a way that this cube will fit into a larger lattice.
For example, a $2^2\times3$ lattice is two cubes side by side but the number of independent gauge links is 5+4=9, not 5+5=10.

\begin{figure}[tb]
\includegraphics[scale=0.5,clip=true,trim=0 0 0 0]{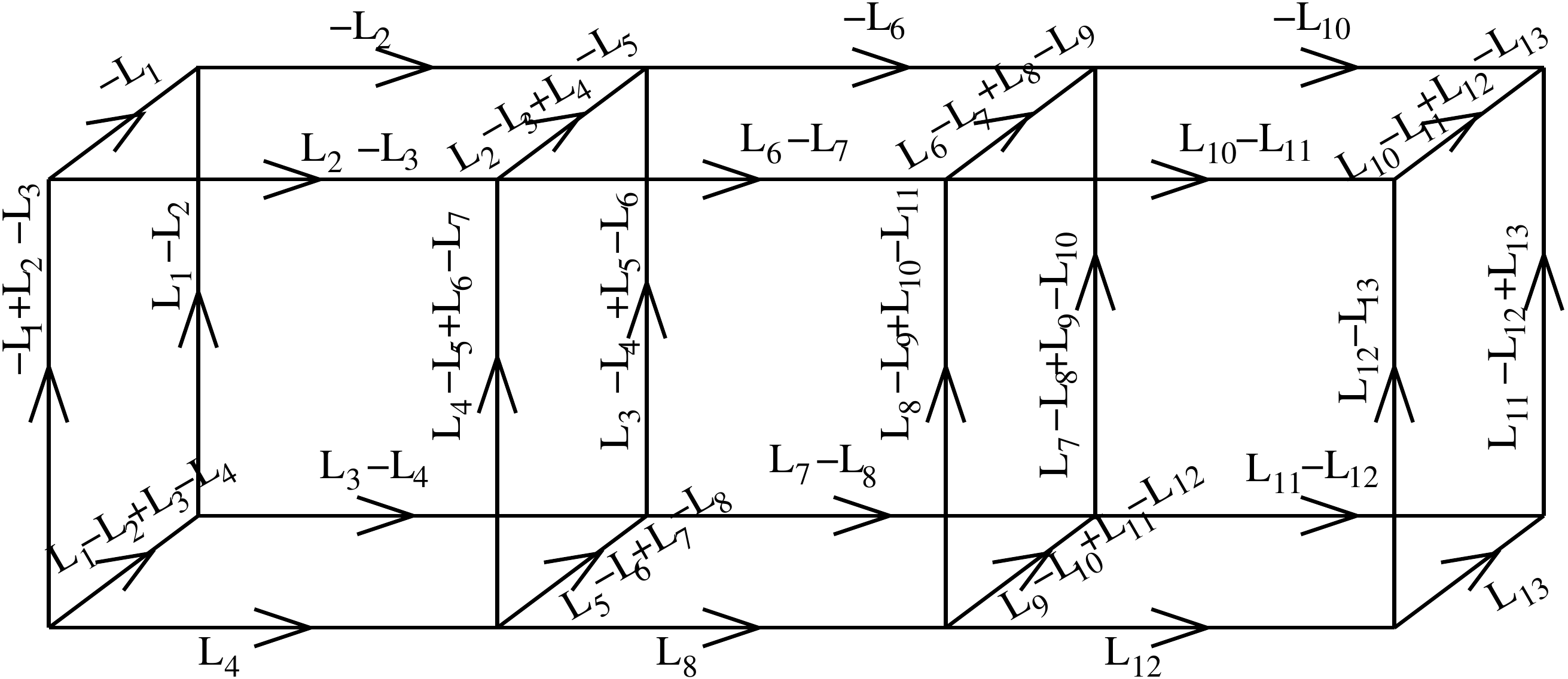}
\caption{A $2\times2\times4$ lattice.
Each gauge link is labeled by its dependence on a minimal set of 13 independent $L_i$ operators.}
\label{fig:2x2xN}
\end{figure}
With $2^2\times N$ lattices in mind, we use the labeling displayed in Fig.~\ref{fig:2x2xN}.
For the first plaquette, notice that $L_1^+$ is a raising operator for the left side,
$L_1^+L_2^+$ is a raising operator for the top,
$L_2^+L_3^+$ is a raising operator for the back,
$L_3^+L_4^+$ is a raising operator for the front,
$L_4^+L_5^+$ is a raising operator for the bottom,
and $L_5^+$ is a raising operator for the right side.
The second cube has $L_5$ already in place for its left side, and adds $L_6$, $L_7$, $L_8$ and $L_9$ as its new links.
Each cube has the same layout of independent links, so the row of cubes can be made arbitrarily long by attaching identical cubes to either end.

From Eq.~(\ref{eq:generalH}), the Hamiltonian for a single cube is
\begin{eqnarray}
H &=& L_1^2 + L_2^2 + L_4^2 + L_5^2 + (L_1-L_2)^2 + (L_2-L_3)^2 + (L_3-L_4)^2 + (L_4-L_5)^2 \nonumber \\
   && + (L_1-L_2+L_3)^2 + (L_3-L_4+L_5)^2 + (L_1-L_2+L_3-L_4)^2 + (L_2-L_3+L_4-L_5)^2 \nonumber \\
   && -\beta^2\left(L_1^+ + L_1^+L_2^+ + L_2^+L_3^+ + L_3^+L_4^+ + L_4^+L_5^+ + L_5^+ + h.c.\right) \nonumber \\
  &=& 4L_1^2 + 6L_2^2 + 6L_3^2 + 6L_4^2 + 4L_5^2 \nonumber \\
   && - 6L_1L_2 + 4L_1L_3 - 2L_1L_4 - 8L_2L_3 + 4L_2L_4 - 8L_3L_4 - 2L_2L_5 + 4L_3L_5 - 6L_4L_5 \nonumber \\
   && -\beta^2\left(L_1^+ + L_1^+L_2^+ + L_2^+L_3^+ + L_3^+L_4^+ + L_4^+L_5^+ + L_5^+ + h.c.\right) \,. \label{eq:cube1}
\end{eqnarray}
Ensuring that no $\beta^2$ term has a product of more than two $L_i^\pm$ operators was one reason for choosing this particular $L_i$ basis.

\begin{figure}[tb]
\includegraphics[scale=0.5,clip=true,trim=0 40 0 80]{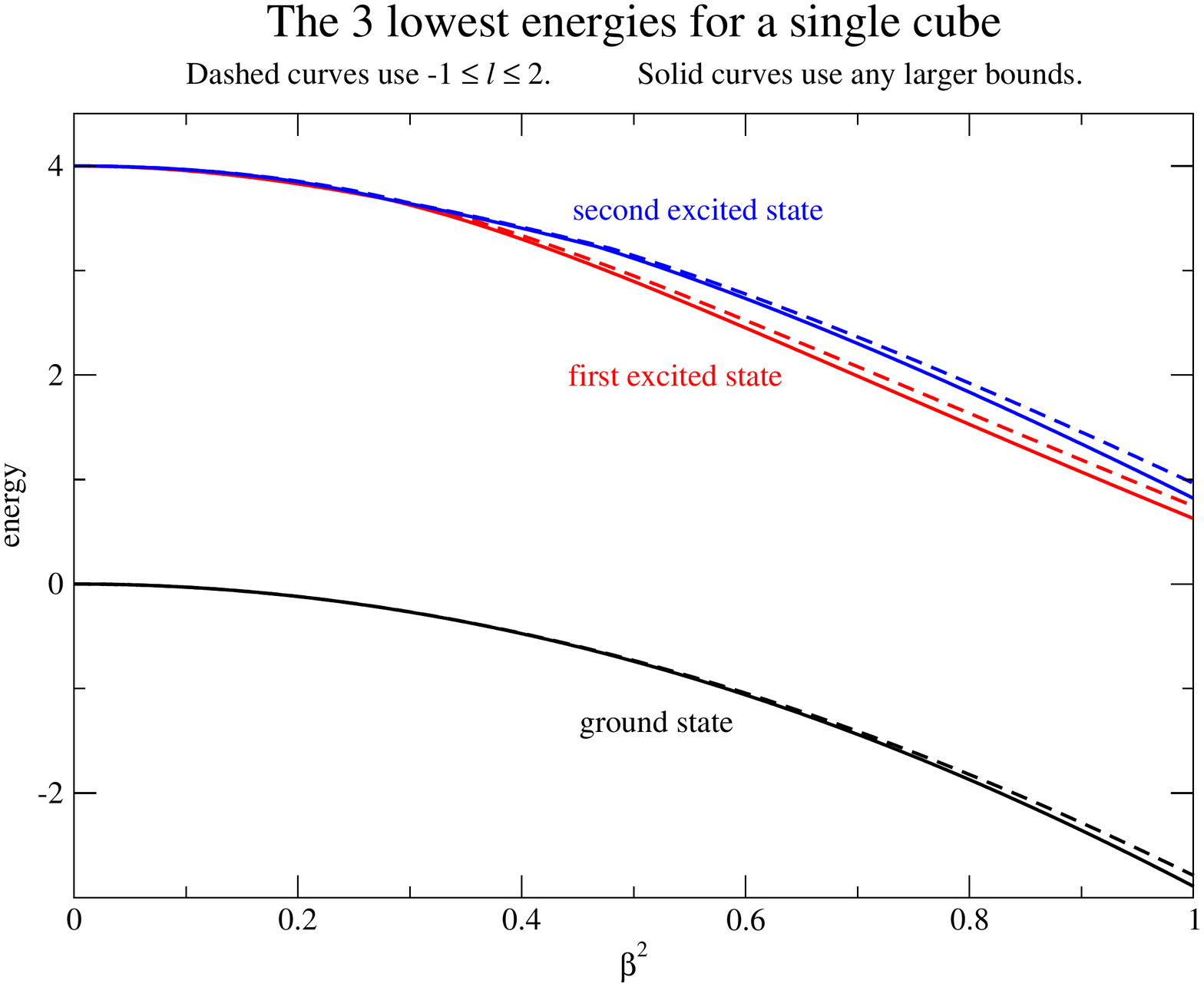}
\caption{The three lowest energy eigenvalues for a single cube (i.e.\ a $2^3$ lattice).
Dashed curves use $-1\leq\ell\leq2$.
Solid curves use any larger bounds.}
\label{fig:cubeLowStates}
\end{figure}

At $\beta^2=0$, the smallest eigenvalue of this Hamiltonian is zero, the first excited state has the value 4 and
is 12-fold degenerate, and the next state has value 6.
Increasing $\beta^2$ breaks the degeneracies.
Choosing two qubits per gauge link gives four basis states per gauge link and makes this Hamiltonian a $2^{10}\times2^{10}$ matrix
which, in contrast to Eq.~(\ref{eq:Hmatrix}), is not tridiagonal.
The effect of truncating to two qubits per gauge link is demonstrated by comparing the dashed curves to the solid curves in Fig.~\ref{fig:cubeLowStates},
which shows the smallest eigenvalues in the strong-coupling region as calculated by matrix diagonalization.

\begin{figure}[tb]
\includegraphics[scale=1.0,clip=true,trim=70 32 90 105]{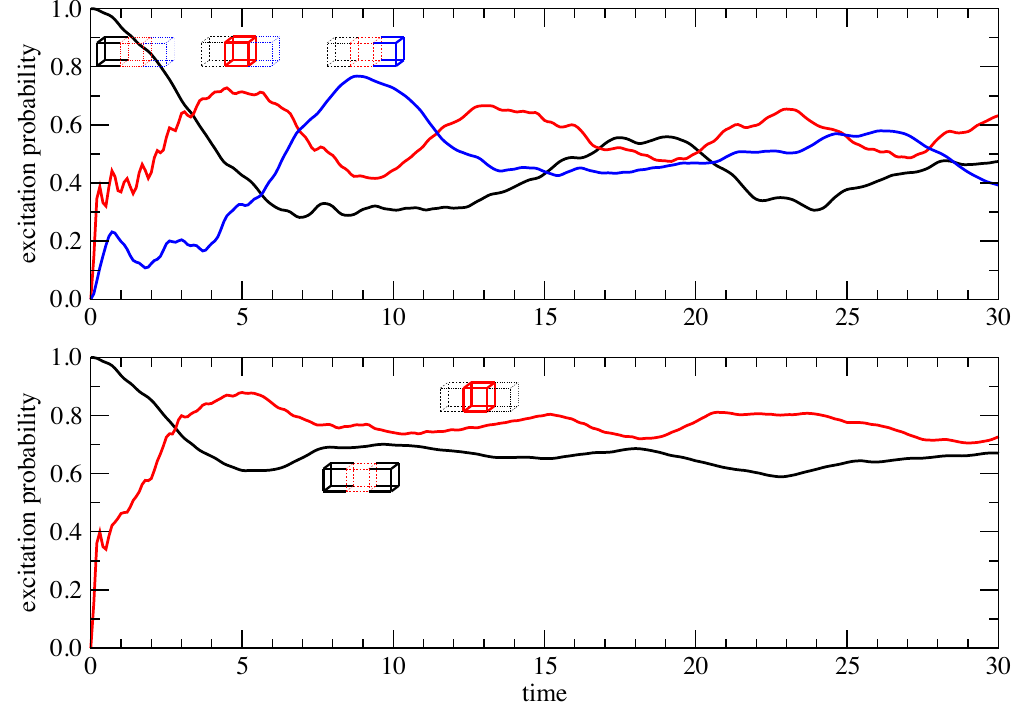}
\caption{Time evolution on a $2^2\times 4$ lattice at $\beta^2=1$ using 2 qubits per gauge link.
The initial state has the cube at one end (upper plot) or the cubes at both ends (lower plot) in the single-cube first excited state and all others in the single-cube ground state.}
\label{fig:cubePropagation}
\end{figure}

To calculate time evolution, we use the three-cube lattice of Fig.~\ref{fig:2x2xN}.
A first thought is to put $L_1$ in the single-plaquette first excited state and all other plaquettes in the single-plaquette ground state.
However, every gauge link on this lattice is shared among multiple plaquettes, and the time evolution does not display clear probability peaks like those observed for the planar lattices in Sec.~\ref{sec:rowplaq}.
Another initial condition is to put the entire left cube into the single-cube first excited state, leaving the other cubes in the single-cube ground state.
That option has gauge links shared between neighbouring cubes, but the upper panel of Fig.~\ref{fig:cubePropagation} does show a traveling excitation for some time until the signal disperses.
The lower panel of Fig.~\ref{fig:cubePropagation} begins with both end cubes in the first excited state and the center cube in the ground state, leading to rapid dispersal of the signal.
A longer lattice and a different choice for the initial state could allow the excitation to persist for a longer time, but we will leave this for future work.

\begin{figure}[tb]
\includegraphics[scale=0.5,clip=true,trim=0 0 0 0]{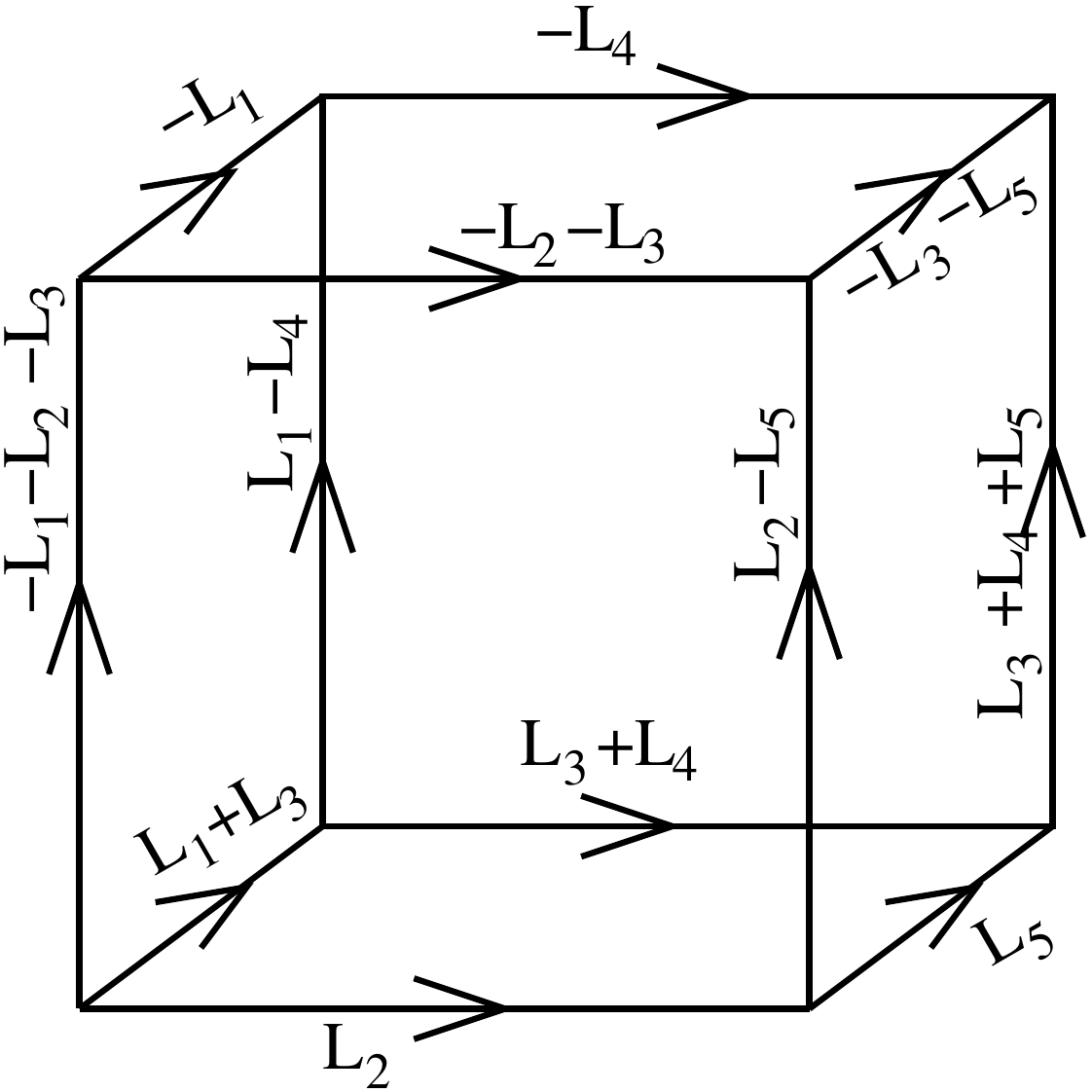}
\caption{A choice for the five independent gauge links within an elementary cube, such that a plane of cubes
can be constructed which satisfies Gauss's law.
Cubes should not be adjoined to the top or bottom of this cube, but can be adjoined to any of the sides.}
\label{fig:altcube}
\end{figure}

Eq.~(\ref{eq:cube1}) is a convenient choice for a row of cubes, but it would not be convenient for a plane of cubes.
The reason is that only two sides of the cube have gauge links ($L_1$ and $L_5$) that are ready to serve a neighbouring cube.
Now we will write down a different labeling for a cube:
\begin{eqnarray}
H &=& L_1^2 + L_2^2 + L_4^2 + L_5^2 + (L_1+L_3)^2 + (L_1-L_4)^2 + (L_2+L_3)^2 + (L_2-L_5)^2 \nonumber \\
   && + (L_3+L_4)^2 + (L_3+L_5)^2 + (L_1+L_2+L_3)^2 + (L_3+L_4+L_5)^2 \nonumber \\
   && - \beta^2(L_1^++L_2^++L_2^+L_3^-L_5^++L_1^+L_3^-L_4^++L_4^++L_5^++h.c.) \nonumber \\
  &=& 4L_1^2 + 4L_2^2 + 6L_3^2 + 4L_4^2 + 4L_5^2 + 2(L_1-L_5)(L_2-L_4) + 4L_3(L_1+L_2+L_4+L_5) \nonumber \\
   && - \beta^2(L_1^++L_2^++L_2^+L_3^-L_5^++L_1^+L_3^-L_4^++L_4^++L_5^++h.c.)
\label{eq:cube2}
\end{eqnarray}
which corresponds to Fig.~\ref{fig:altcube}.
Cubes of this type can immediately be used side by side to fill a plane, because $L_1^+$, $L_2^+$, $L_4^+$ and $L_5^+$ are
the raising operators for the four sides of the cube and are directly usable by neighbouring cubes.
However, Eq.~(\ref{eq:cube2}) requires significantly more Pauli factors than Eq.~(\ref{eq:cube1}) due to the triple products of $L_i^\pm$ operators in the terms proportional to $\beta^2$, so there is a price to be paid when moving beyond a row of cubes.

\begin{figure}[tb]
\includegraphics[scale=0.5,clip=true,trim=0 0 0 0]{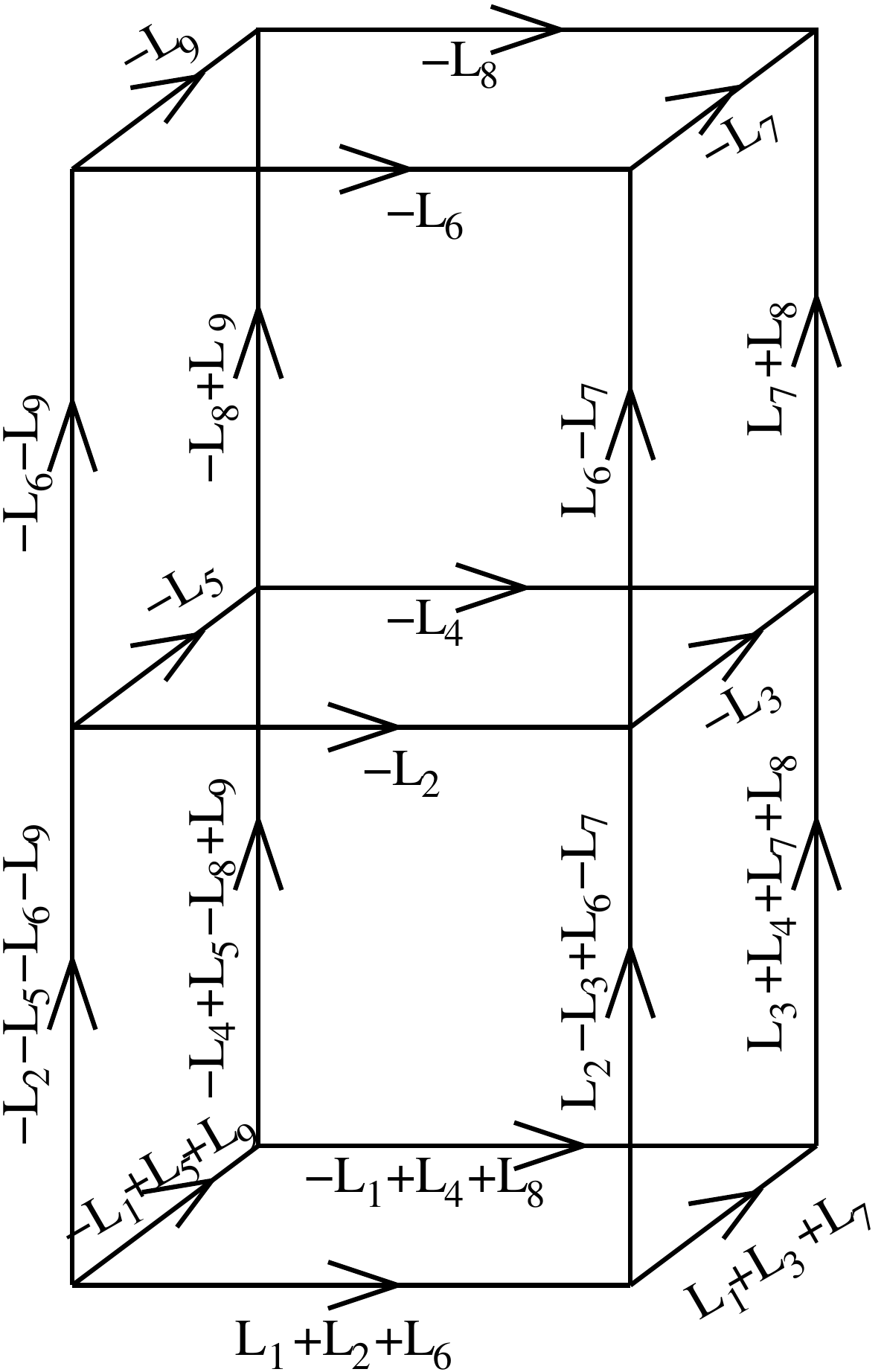}
\caption{The template for a lattice tower that can be extended to any height and any depth.
It can immediately be adjoined to other towers on all sides to form a lattice of any volume, while maintaining Gauss's law.}
\label{fig:volume}
\end{figure}

A final step is to extend the lattice into a 3-dimensional volume.
This is necessarily more expensive than for a row or plane of cubes, but there is a recommendation that maintains Gauss's law
and requires no more than five $L_i^\pm$ operators in a single $\beta^2$ term.
This remains true for any size of the lattice volume.
The recommendation is to build cube towers of any height following the template in Fig.~\ref{fig:volume}, where the floor's raising operator is $L_1^+$ and every wall plaquette is a product of one or two $L_i^\pm$ factors.
Towers like this can immediately be put side by side to create any lattice volume.
The Hamiltonian for the two-cube example in Fig.~\ref{fig:volume} is
\begin{eqnarray}
H &=& (L_1+L_2+L_6)^2 + (L_1+L_3+L_7)^2 + (L_1-L_4-L_8)^2 + (L_1-L_5-L_9)^2 \nonumber \\
   && + (L_2-L_3+L_6-L_7)^2 + (L_3+L_4+L_7+L_8)^2 + (L_2+L_5+L_6+L_9)^2 \nonumber \\
   && + (L_4-L_5+L_8-L_9)^2 + L_2^2 + L_3^2 + L_4^2 + L_5^2 + L_6^2 + L_7^2 + L_8^2 + L_9^2 \nonumber \\
   && + (L_6-L_9)^2 + (L_7+L_8)^2 + (L_8-L_9)^2 + (L_6+L_9)^2 \nonumber \\
   && - \beta^2(L_1^+ + L_2^+ + L_3^+ + L_4^+ + L_5^+ + L_1^-L_2^+L_3^+L_4^+L_5^+ \nonumber \\
   &&           +L_2^-L_6^+ + L_3^-L_7^+ + L_4^-L_8^+ + L_5^-L_9^+ + L_1^-L_6^+L_7^+L_8^+L_9^+ + h.c.) \,.
\end{eqnarray}
To add another cube on top of the tower shown in Fig.~\ref{fig:volume}, we would define $L_{10}$ around the entire 8-link path on the front of the tower, just as $L_6$ appears in the entire path on the front of the present tower.
Similar definitions apply to $L_{11}$, $L_{12}$ and $L_{13}$ on the other sides of the tower.
An ever higher tower requires ever longer paths for the newest gauge links, representing a nonlocality in the Hamiltonian that
was not required for a row or plane of plaquettes.

\section{Conclusions}\label{sec:concl}

In this work, we have presented a model for compact U(1) lattice gauge theory in 2 or 3 spatial dimensions.
Gauss's law is implemented fully, leaving no redundant gauge field degrees of freedom.
The model can be implemented with any number of qubits per gauge link.
The smallest eigenvalues of the single plaquette and the single cube, plotted as functions of $\beta^2$ in Figs.~\ref{fig:plaqLowStates} and \ref{fig:cubeLowStates} respectively, indicate that a truncation to just two qubits per gauge link is already a useful model.

This work reports on calculations performed on $2\times N$ lattices, i.e.\ a row of plaquettes, showing that simple initial conditions produce visible propagation and collisions of excitations.
Calculations on nine plaquettes forming a square lattice produce similar results, though excitations disperse more quickly when this extra spatial dimension is available.
Extending the model to a 3-dimensional volume requires a non-locality that was not present in the planar Hamiltonian, and an implementation of the 3D model has been presented in this work.

U(1) is the simplest gauge theory, and the qubit model presented here is useful even with an aggressive truncation.
Therefore this model provides an opportunity to explore many aspects of quantum computation using only a small number of qubits.
An interesting next step could be to design initial states that allow excitations to propagate for longer times on planar or 3D lattices.
We envision such studies being performed by simulators running on classical computers, as was done in the present work, but perhaps it will not be too long until the model can be implemented directly on quantum computers as well.

\section*{\hspace{-6.5mm} Acknowledgments}

RL is grateful to the organizers of {\em Lattice for BSM Physics 2019} 
for the opportunity to present this work and to the participants for valuable conversations.
This work was supported in part by the Natural Sciences and Engineering Research Council (NSERC) of Canada.
TRIUMF receives federal funding via a contribution agreement with the National Research Council (NRC) of Canada.

\end{document}